\documentstyle[psfig, epsf]{mn}

\newif\ifAMStwofonts

\newcommand{\Cpp}{C\raisebox{0.45ex}{\scriptsize++}\normalsize}
\newcommand{\mcII}{MOA-cam2}
\newcommand{\re}{r_{\rm E}}
\newcommand{\umin}{u_{\rm min}}
\newcommand{\that}{t_{\rm E}}
\newcommand{\amax}{A_{\rm max}}
\newcommand{\tmax}{t_{\rm max}}
\newcommand{\fr}{F_{\rm ref}}
\newcommand{\fb}{F_{\rm base}}
\newcommand{\chisq}{\chi^2}
\newcommand{\vbar}{\tilde{v}}

\ifoldfss
  \ifCUPmtlplainloaded \else
    \NewTextAlphabet{textbfit} {cmbxti10} {}
    \NewTextAlphabet{textbfss} {cmssbx10} {}
    \NewMathAlphabet{mathbfit} {cmbxti10} {} 
    \NewMathAlphabet{mathbfss} {cmssbx10} {} 
  \fi
  \ifAMStwofonts
    \ifCUPmtlplainloaded \else
      \NewSymbolFont{upmath} {eurm10}
      \NewSymbolFont{AMSa} {msam10}
      \NewMathSymbol{\upi}     {0}{upmath}{19}
      \NewMathSymbol{\umu}     {0}{upmath}{16}
      \NewMathSymbol{\upartial}{0}{upmath}{40}
      \NewMathSymbol{\leqslant}{3}{AMSa}{36}
      \NewMathSymbol{\geqslant}{3}{AMSa}{3E}

    \fi
  \fi
\fi 

\ifnfssone
  \newmathalphabet{\mathit}
  \addtoversion{normal}{\mathit}{cmr}{m}{it}
  \addtoversion{bold}{\mathit}{cmr}{bx}{it}
  \newmathalphabet{\mathbfit} 
  \addtoversion{normal}{\mathbfit}{cmr}{bx}{it}
  \addtoversion{bold}{\mathbfit}{cmr}{bx}{it}
  \newmathalphabet{\mathbfss} 
  \addtoversion{normal}{\mathbfss}{cmss}{bx}{n}
  \addtoversion{bold}{\mathbfss}{cmss}{bx}{n}
  \ifAMStwofonts
    \ifCUPmtlplainloaded \else
      \UseAMStwoboldmath
      \makeatletter
      \new@mathgroup\upmath@group
      \define@mathgroup\mv@normal\upmath@group{eur}{m}{n}
      \define@mathgroup\mv@bold\upmath@group{eur}{b}{n}
      \edef\UPM{\hexnumber\upmath@group}
      \new@mathgroup\amsa@group
      \define@mathgroup\mv@normal\amsa@group{msa}{m}{n}
      \define@mathgroup\mv@bold\amsa@group{msa}{m}{n}
      \edef\AMSa{\hexnumber\amsa@group}
      \makeatother
      \mathchardef\upi="0\UPM19
      \mathchardef\umu="0\UPM16
      \mathchardef\upartial="0\UPM40
      \mathchardef\leqslant="3\AMSa36
      \mathchardef\geqslant="3\AMSa3E
    \fi
  \fi
\fi 

\ifnfsstwo
  \DeclareMathAlphabet{\mathbfit}{OT1}{cmr}{bx}{it}
  \SetMathAlphabet\mathbfit{bold}{OT1}{cmr}{bx}{it}
  \DeclareMathAlphabet{\mathbfss}{OT1}{cmss}{bx}{n}
  \SetMathAlphabet\mathbfss{bold}{OT1}{cmss}{bx}{n}
  \ifAMStwofonts
    \ifCUPmtlplainloaded \else
      \DeclareSymbolFont{UPM}{U}{eur}{m}{n}
      \SetSymbolFont{UPM}{bold}{U}{eur}{b}{n}
      \DeclareSymbolFont{AMSa}{U}{msa}{m}{n}
      \DeclareMathSymbol{\upi}{0}{UPM}{"19}
      \DeclareMathSymbol{\umu}{0}{UPM}{"16}
      \DeclareMathSymbol{\upartial}{0}{UPM}{"40}
      \DeclareMathSymbol{\leqslant}{3}{AMSa}{"36}
      \DeclareMathSymbol{\geqslant}{3}{AMSa}{"3E}
    \fi
  \fi
\fi 

\ifCUPmtlplainloaded \else
  \ifAMStwofonts \else 
    \def\upi{\pi}
    \def\umu{\mu}
    \def\upartial{\partial}
  \fi
\fi

\title{Real-Time Difference Imaging Analysis of MOA Galactic Bulge Observations
During 2000}
\author[I.A. Bond et al.]{
I.A.~Bond,$^{1,2}$ F.~Abe,$^3$ R.J.~Dodd,$^{1,4,5}$
J.B.~Hearnshaw,$^2$ M.~Honda,$^6$ 
\newauthor J.~Jugaku,$^7$ P.M.~Kilmartin,$^{1,2}$ A. Marles,$^1$ 
K.~Masuda,$^3$ Y.~Matsubara,$^3$ 
\newauthor Y.~Muraki,$^3$ T.~Nakamura,$^8$ G.~Nankivell,$^5$ S.~Noda,$^3$ 
C.~Noguchi,$^3$  
\newauthor K.~Ohnishi,$^9$ N.J.~Rattenbury,$^1$ M.~Reid,$^4$ To.~Saito,$^{10}$ 
H.~Sato,$^8$ 
\newauthor M.~Sekiguchi,$^6$ J.~Skuljan,$^2$
D.J.~Sullivan,$^4$ T.~Sumi,$^3$ M.~Takeuti,$^{11}$ 
\newauthor Y.~Watase,$^{12}$ S.~Wilkinson,$^4$ R.~Yamada,$^3$ 
T.~Yanagisawa,$^3$ 
\newauthor and P.C.M.~Yock$^1$
\\ 
$^1$Faculty of Science, University of Auckland, Auckland, New Zealand\\ 
$^2$Department of Physics and Astronomy, University of Canterbury, 
Christchurch, New Zealand\\ 
$^3$Solar-Terrestrial Environment Laboratory, Nagoya University, Nagoya 464, 
Japan\\ 
$^4$School of Chemical and Physical Sciences, Victoria University, 
Wellington, New Zealand\\ 
$^5$Carter Observatory, P.O. Box 2909, Wellington, New Zealand\\ 
$^6$Institute of Cosmic Ray Research, University of Tokyo, Tanashi, Tokyo 188, Japan\\ 
$^7$Research Institute of Civilization, Tama 206, Japan\\
$^8$Department of Physics, Kyoto University, Kyoto 606, Japan\\
$^9$Nagano National College of Technology, Japan\\ 
$^{10}$Tokyo Metropolitan College of Aeronautics, Tokyo 116, Japan\\ 
$^{11}$Tohoku University, Sendai, Japan\\
$^{12}$KEK Laboratory, Tsukuba 305, Japan
}

\date{Accepted .
      Received ;
      in original form }

\pagerange{\pageref{firstpage}--\pageref{lastpage}}
\pubyear{1994}

\begin{document}

\maketitle

\label{firstpage}

\begin{abstract}
We describe observations carried out by the MOA group of the Galactic
Bulge during 2000 that were designed to detect efficiently gravitational
microlensing of faint stars in which the magnification is high and/or
of short duration. These events are particularly useful for studies of
extra-solar planets and faint stars. Approximately 17 deg$^2$ were
monitored at a sampling rate of up to 6 times per night. The images were
analysed in real-time using a difference imaging technique. Twenty microlensing
candidates were detected, of which 8 were alerted to the microlensing
community whilst in progress. Approximately half of the candidates had
high magnifications ($\ga10$), at least one had very high magnification
($\ga50$), and one exhibited a clear parallax effect. The details of these 
events are reported here, together with details of the on-line difference imaging 
technique. Some nova-like events were also observed and these are described,
together with one asteroid.
\end{abstract}


\section{Introduction}

The observation of gravitational microlensing events is becoming established 
as an important tool in several astrophysical endeavours. These
include the search for extra-solar planets, studies of dark matter,
studies of stellar atmospheres, 
and studies of galactic structure. More than 500 microlensing events,
most of them alerted in real time, have been identified by the
microlensing survey groups MACHO \cite{al93}, OGLE \cite{ul92}, 
and EROS \cite{aub}. These groups monitored some millions of stars within
their fields of view to find the microlensing events. The photometry 
measurements were based on
profile fitting software such as Dophot \cite{sch} which 
determines fluxes by integrating over the profile of a given star on 
a given image.

While profile fitting techniques have proved to be highly successful, the
limitations of this type of analysis have been recognized for some time.
The most serious limitation is the contamination of flux measurements 
by the ``blending'' of the profiles of neighbouring stars. This problem 
becomes more serious when one encounters crowded regions of stars within
the fields of view. Blending causes systematic errors in photometry
measurements and subsequent systematic errors in the determination
of microlensing parameters such as event duration and peak 
amplification \cite{han}. Alard \shortcite{al01} has shown that, 
using an image subtraction or difference imaging analysis (hereafter DIA)
to allow for the effects of blending, an
unbiased reconstruction of the mass function in the Galactic bulge
should be possible using microlensing data alone.

Another problem with profile fitting analyses
is that only those microlensing events whose source stars are visible
and resolvable at baseline are detected. Microlensing events that
rise from below the observational threshold to be visible during times
of high magnification are missed. This is unfortunate, because these 
events with faint source stars are potentially capable of providing 
valuable information. The mere detection of faint stars in the Galactic
Bulge that are normally undetectable can clearly provide useful information.
For example, if the magnification is
high, spectroscopic information may also be obtained that is not 
available otherwise (Lennon et al. 1996, 1997; Minniti et al.1998). 
Kane and Sahu (2001) recently made a kinematic study of the far side
of the Galactic bulge using spectroscopy of microlensed sources.

The peaks of high magnification events provide valuable information
on extra-solar planetary systems. Griest and Safizadeh (1998) showed
that the probability for detecting Jupiter-like planets is nearly
100\% for events with peak amplifications $\ga100$ provided good 
photometric sampling is carried out around the times of the peaks,
and that the probability for detecting Earth-like planets is substantial.
They also showed the probability for detecting Jupiter-like planets is
substantial in events with peak amplifications $\ga10$. These predictions
have been borne out by observation (Rhie et al. 2000; Albrow et al. 2000;
Bond et al. 2001). As will be shown here, high magnification events can be
detected efficiently using DIA.

DIA largely
overcomes the problems caused by blended or undetected microlensing
source stars. Recent re-analyses of the MACHO database using DIA have 
increased the microlensing detection rate and improved the 
quality of the photometry (Alcock et al. 1999a; 1999b; 2000). A re-analysis
of the OGLE database is underway and similar results are emerging 
\cite{woz}. All of these analyses were carried out off-line. 

During 2000 we applied DIA to observations of the Galactic Bulge carried
out by the MOA Group. The analysis was ``real-time'' in that alerts of
microlensing events in progress were provided to the microlensing community.
In this paper we describe our implementation of the real-time DIA method and 
present the transient phenomena observed during 2000. 

\section{Observations by MOA of the Galactic Bulge}

The observations were carried out using the 60-cm Boller \& Chivens
telescope of the Mt John University Observatory in New Zealand. The 
telescope was modified by the MOA group to provide a 1\fdg3 field of
view at f/6.25 and to enable computer controlled tracking (Abe et al. 1997).
The data were taken using the wide
field camera, MOA-cam2, that was constructed by the MOA group \cite{yan}. 
MOA-cam2 is a mosaic camera comprised of three abutted thinned 
2k$\times$4k SITe
CCD chips. Two wide passbands were employed: a 
``blue'' (400--630 nm) and a ``red'' (630--1000 nm) band. The wide passbands
compensate for the small telescope aperture. The total field of view over
the three CCDs is $55\arcmin\times83\arcmin$ and the pixel size on the sky is 
$0.81\arcsec$. This is matched to the less than ideal seeing of the set-up
which ranges from about $1.8\arcsec$ to $3.5\arcsec$, with the median at 
about $2.5\arcsec$.
These are global figures that include dome seeing, etc.

The observational strategy employed by MOA for the Galactic Bulge in 2000 was
somewhat different from that of the established survey groups MACHO, OGLE, and
EROS. Rather than sample a large area of the Bulge once per night, we chose
to sample a smaller area several times per night. This decision was taken 
mainly with the aim of improving our sensitivity to very high magnification
events which are expected
to rise and fall very rapidly and only be detected above the observation 
threshold for a few days or less.

For the 2000 observations, a total of 14 \mcII~fields towards the 
Galactic Bulge were selected on the basis of the distribution of 
microlensing events detected by the MACHO Collaboration during 1995--1999. 
\footnote{These observation fields are a revision of previously defined 
survey fields observed with \mcII~ during 1998--99. The
data from observation fields used prior to 2000 were not used in the
survey analysis in the present study, but they were used for follow-up
analysis of selected events that are presented below.}
These fields cover an area of $\sim17$ deg$^2$. The exposure time was 
180 seconds per field
with most exposures being taken in the red passband. Exposures in the
blue passband were only made occasionally and are not used in this study. 
This observational strategy
allowed a sampling rate for the full 14 fields of up to 6 times per night.
These fields were observed by MOA between 2000 March and September.
Problems with the camera electronics prevented observations over a six-week 
period during July--August.

\section{Data Analysis}

\subsection{Image Subtraction Method}

\begin{figure*}
   \caption{Difference imaging in action. The top left image is a sub-region
      of a reference image. The top right image is a corresponding
      observation image. The squares in the two top images mark the location 
      of stamps used to calculate the kernel. The lower left image is
      the result of subtracting
      the reference image from the observation image. The lower right 
      image is the same subtracted image with the positions of saturated
      stars masked out.}
   \label{diffs}
\end{figure*}

Two geometrically aligned images taken under conditions of different seeing 
are related through the convolution relation

\begin{equation}
i(x, y) = r * k (x, y) + b(x, y).
\end{equation}
Here $r$ is the better seeing image referred to as the ``reference'' image
and $i$ is the current ``observation'' image. The convolution kernel $k$
encodes the seeing differences and $b$ represents the sky background
differences between the two images. Solving for the convolution
kernel is the crucial step in image subtraction. There are two approaches
in use for doing this. One matches point spread
function (PSF) models on the two images by solving for the kernel in
Fourier space (Phillips \& Davis 1995; Tomaney \& Crotts 1996). 
This approach was adopted by Alcock et al. \shortcite{al99a} 
in their re-analysis of a subset of the MACHO data.

The other 
approach is to directly model the kernel in real space \cite{alu}. 
We developed our own implementation of this method.
This includes the modification of Alard \shortcite{al} that models
spatial variations of the kernel across the CCD. The mathematical
techniques are described fully by Alard \& Lupton \shortcite{alu} and 
Alard \shortcite{al}. Here we summarize our implementation. The 
convolution kernel function at a given location $(x_0, y_0)$
is expressed as a linear combination of analytical ``basis'' functions

\begin{equation}
k(x_0, y_0, x, y) = a_0 f_0(x, y) + \sum_m a_m(x_0,y_0) f_m(x,y).
\end{equation}
The coefficients in the linear combination are spatially dependent.
They are also modeled as a linear combination of basis functions which takes
the form of a two dimensional polynomial

\begin{equation}
a_m(x_0,y_0) = \sum_n a_{mn} x_0^p y_0^q,
\end{equation}
where the polynomial indices $p$ and $q$ depend implicitly on the function
index $n$. The basis functions used to model the kernel take the form
of a combination of two dimensional Gaussian functions and polynomials. The
function $f_m(x,y)$ in Eqn.~2 can be expressed in such a way as to render the
integral of each but the first basis function zero. Consequently the integral of the 
kernel itself 
is constant across the chip regardless of the spatial variations of the
kernel. The differential background is modeled in the same 
way as the spatial variations in the kernel:

\begin{equation}
b(x,y) = \sum_n b_n x^p y^q
\end{equation}
where again the polynomial indices depend implicitly on the function index.

The solution for the coefficients ${a_{mn}}$ and ${b_n}$ may be found using 
standard linear techniques. The linear system is constructed using a number
of sub-regions, referred to as ``stamps'', centred on bright, but 
not saturated, stars throughout the
image. An advantage of this method of solution is that these
stamps need not be centred on isolated stars with clear point spread
functions (PSF stars). There can be any number of
stars to any degree of crowding within a stamp. However, we impose the
strict requirement that regions enclosed by the stamps, plus an additional
margin given by half the kernel size\footnote{We used a kernel of dimensions 
$17\times17$ pixels in this analysis.}, are completely free of bad pixels 
including dead columns and bleeding columns due to nearby saturated stars.
Imposing this condition greatly improves the robustness in obtaining
good quality subtracted images. Another consideration is that the size of
the linear system is large when one includes the spatial variations in the
convolution kernel. The computing time required to calculate the linear system
can be greatly reduced using an accelerated summation procedure which assumes
that the kernel is constant within individual stamps \cite{al}.

Our implementation of the kernel solution method along with the subsequent
image subtraction is coded in \Cpp. It is worth mentioning here the merits of
using an object oriented programming language for this purpose.
Eqns.~2--4 are somewhat unwieldy to encode in a procedurally oriented
language such as Fortran or C. The object oriented feature of
\Cpp~allows one to encapsulate neatly a set of basis functions and their
associated polynomial index tables in a single class. One then declares
separate class instances for each set of basis functions describing the
kernel, its spatial variations and the differential background.

The PSF profiles are known to vary across the field for large CCD images.
Consequently, the possible spatial variation of the convolution kernel is an
important consideration. 
The approach adopted by the MACHO collaboration
was to divide their images into small sub-rasters and solve for the
convolution kernel separately on each of these sub-rasters assuming
no significant variations in the kernel occurred on their length 
scales \cite{al99a}. However, for the MOA images, we find small
but significant spatial variations of the kernel even over
small $500\times500$ sub-regions of the field. This
prompted us to select the direct kernel modeling method which can 
easily model these spatial variations.

Another consideration which influenced our choice of kernel solution 
method was that
the PSF matching method requires high signal-to-noise empirical PSF
models on both the reference and observation images. This would be 
difficult to achieve consistently using a small telescope such as
ours at a site with only moderately good seeing. On the other hand, the
direct kernel modeling method only requires the reference image to be
of high signal-to-noise.

An example of a subtracted image is shown in Fig.~\ref{diffs}.
The approach taken to subtract a MOA 2k$\times$4k observation image
from its corresponding reference image involved first selecting suitable
stamps on the reference image.  Typically $\sim400$ stamps
were selected per 2k$\times$4k field. An observation and reference
image were then tiled into 1k$\times$1k sub-images which were small
enough to allow the spatial variations in the kernel and differential
sky background to be satisfactorily modeled using smooth functions.
The solution for the convolution kernel and its associated spatial 
variations, together with the differential sky backgrounds, was computed
for each sub-region. 
The corresponding subtracted image was then formed using
\begin{equation}
   \Delta i(x, y) = i(x,y) - r*k(x,y) - b(x,y)
\end{equation}
For a time series of observation images of a given field, a reference
image was formed from the best seeing image or a combination of the
best seeing images. The reference image is fixed over the entire time 
series. The stamps used for the kernel solution were also fixed for the
time series of observations.

A mask image mapping the distribution of saturated stars was also formed 
from the reference image. If a particular star contained
saturated pixels, the entire star profile was masked out. There is
little useful data that can be obtained at the position of saturated
stars. The mask was applied to all subtracted images to 
obtain final images.

\begin{figure*}
   \epsfxsize=\hsize\epsfbox{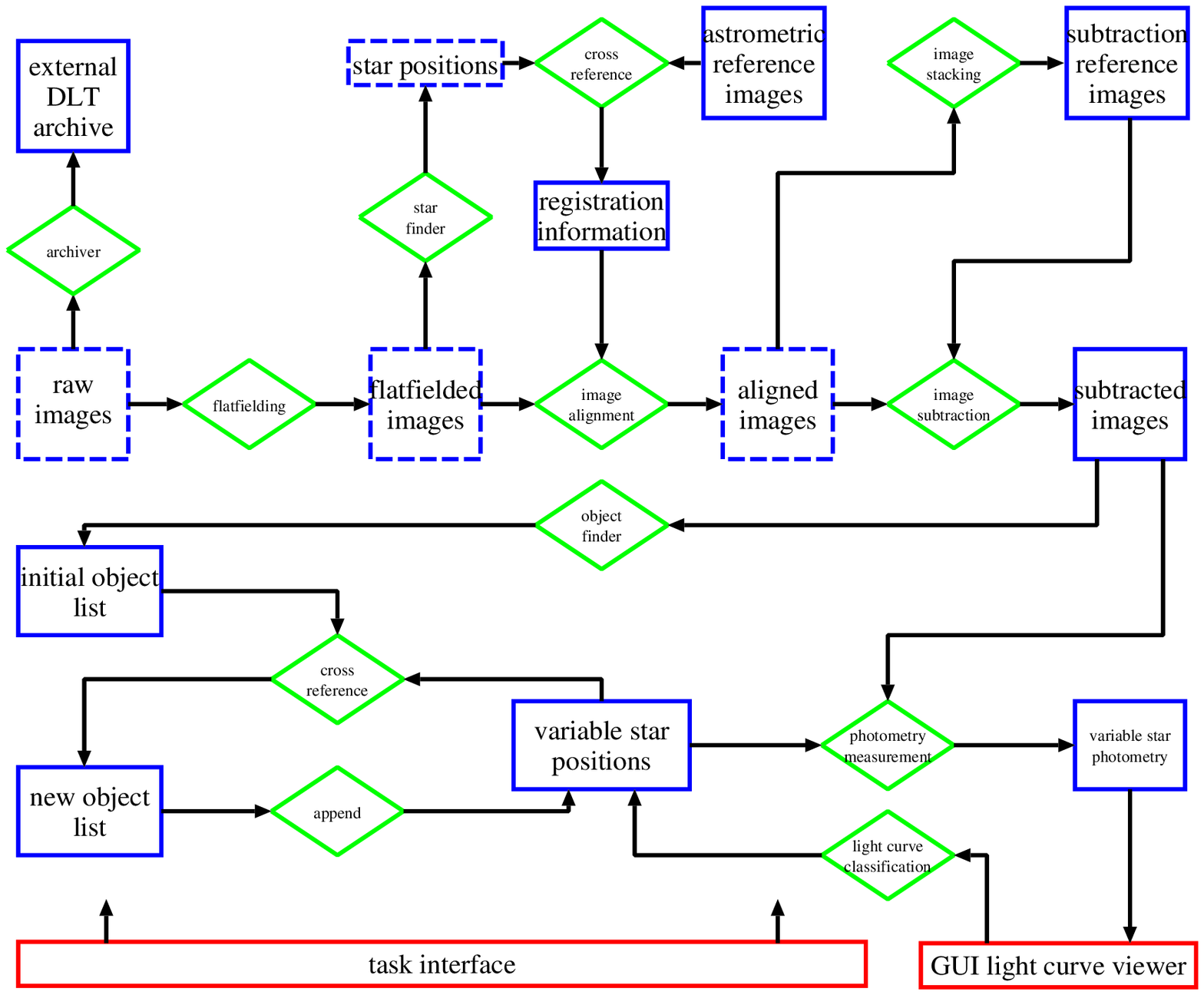}
   \caption{Road map of the on-line data analysis environment. The
	boxes represent the various databases. The dashed boxes
	datasets which are flushed periodically to relieve pressure on
	system disk space. The diamonds represent the analysis
	procedures connecting the databases.}
   \label{roadmap}
\end{figure*}

\subsection{Data Analysis Environment}

In this work, two distinct modes of analysis were employed. The ``online''
analysis mode was used for real time reductions the large quantity
of 2K$\times$4K images collected during observations. A more detailed
and rigorous ``follow-up'' analysis was then applied to individual events
found in the online analysis. The online analysis system
is shown schematically in Fig.~\ref{roadmap}. The process involves
the interplay of a large number of image and numerical datasets. It is
more helpful to regard the system as a data and image management system
rather than a linear data reduction ``pipeline''. The key features of
the analysis system are as follows.

\subsubsection{Task Interface}

Each of the tasks linking the databases shown in Fig.~\ref{roadmap}
is controlled
by a single Perl script. These scripts function as wrappers to lower level
image and data analysis programs most of which are coded in \Cpp. Each task
can be run individually or combined in other scripts to provide a higher
layer of automation. This design provides considerable flexibility in 
operation, ranging from complete automation down to image-by-image 
interaction.

\subsubsection{Image Registration}

The image registration database provides information on the geometric
alignment of observation images to their corresponding astrometric
reference images. The registration process involves cross-referencing
resolved stars on the two images to achieve alignment. Each of the 
2k$\times$4k observation images was sub-divided into 1k$\times$1k 
sub-frames. The 
brightest 1000 stars within each sub-region were found using our
implementation of the algorithm in the IRAF task DAOFIND. The 
star positions in each sub-frame were cross-referenced with star
positions on the corresponding sub-frame of the astrometric reference
using a smart algorithm which starts with the pre-supposition that the
brighter stars on the observation image will be correspondingly bright
on the astrometric reference image. A slower but very robust
alternative algorithm was used as a backup if necessary. This combination 
of algorithms provided an extremely robust frame-to-frame matching procedure
that worked for all images taken under a variety of viewing conditions.
For each observation image one finally obtained
a list of 6000--8000 cross referenced stars over the full 2k$\times$4k
field. The positions of these stars on the observation and reference
images were entered into the registration database.

\subsubsection{Image Alignment}

Geometrically aligned images were formed from flat-fielded images
using information in the registration database. The alignment 
task was controlled by a Perl wrapper script for the IRAF tasks GEOMAP
and GEOTRAN. This alignment process was used in the construction of 
subtraction reference images and in preparing images for the
subtraction process itself.

\subsubsection{Reference Image Databases}

These databases contain the astrometric and image subtraction reference
images for each CCD of each observation field. The astrometric reference 
images define the coordinate system to which all observation images were 
geometrically aligned. The images used for both astrometry and subtraction 
were selected from amongst the best seeing, highest signal to noise, and 
lowest airmass images obtained early in the 2000 Galactic Bulge observation 
season.We decided not to form subtraction reference images by combining several
images. Instead we chose just one good seeing image per field as the reference.
When combining large images such as ours, there is a danger of corrupting the
linearity in the final image due to non-uniformities in the sky background across
the chip. In image subtraction, it is crucial that the pixel values associated
with the images are linear with respect to the incident flux..

\subsubsection{From Raw to Subtracted Images}

The backbone of the image reduction flow involves processes
that generate subtracted images from raw observation images. 
Intermediate image databases contain
flat-fielded and aligned images. To relieve pressure on computer
disk space, raw, flat-fielded, and aligned image databases
were flushed periodically. However, all raw images were externally
archived onto DLT tapes. When required, a given set of flat-fielded
and aligned images could easily be regenerated using information in
the registration database. On the other hand, all subtracted images
were retained on disk as these were constantly being accessed 
by the photometry measurement processes.

\subsubsection{Identifying Objects on Subtracted Images}

The positions of variable stars on a given subtracted image exhibited
positive or negative profiles depending on whether the flux had increased
or decreased relative to the reference image. The subtracted images
also contained spurious profiles not associated with stellar variability.
One type of spurious profile arose from the effects of differential 
refraction. Dependent upon the star's colour and the airmass at the time of 
observation, the position of its centroid was sometimes displaced
to a greater extent than other stars. The result
was an imperfect subtraction at the star's position, giving a combined positive
and negative profile. We did not attempt to correct for this effect, but
instead filtered out these imperfect subtractions (see Alcock et al. 1999b
for a discussion on how to deal with differential refraction).

The first step in identifying objects associated with genuine variability
was the application of the star finding algorithm described previously to the
subtracted image to produce an initial list of variable star positions.
This list was then screened by applying a second algorithm which examined the
profile at each position. Essentially, if the profile contained a large
number of either positive definite {\it or} negative definite pixel values,
then the profile was deemed to be that of a genuine variable star. For each
CCD on each field, newly detected objects were assigned running numbers 
according to discovery sequence.

\subsubsection{Photometry Databasing}

We did not maintain a database of light curves of all resolved stars across
the observation fields. Rather, we maintained a database of variable star 
positions coupled with the database of subtracted images. Light curves
were calculated only for objects of ``current interest'' such as newly
identified transient events. The variable star position database was updated 
periodically as new variables were identified and more refined positions were
calculated. Since the subtracted images were retained on disk, a light
curve at any given set of positions could be easily generated. In the online 
analysis, we extracted photometric measurements from the subtracted images
using aperture photometry with an aperture radius of 6 pixels. In the follow-up
analysis described in the next section, we used more a precise measurement
technique.

\subsubsection{Event Selection}

We wished to detect new transient events online as opposed to periodic
variables such as Miras, RR Lyrae stars, etc. A search for new events was
generally carried out at the end of each night of good seeing conditions. 
For each
field, the sequence of subtracted images from the previous night 
was combined to form a 
mean subtracted image which was examined for positive and negative profiles
due to stellar variability. The positions of these profiles were
cross-referenced with those of variable stars previously detected to
produce a list of ``new'' variables. The variable star database was then
updated by appending the new detections. Light curves for the new detections
were generated by performing aperture photometry at their respective 
positions on the entire sequence of subtracted images up to the current
observation date. This practice of periodically carrying out new searches for
events rendered the procedure sensitive to short timescale transient events. 

The light curves corresponding to all new detections were scrutinized by eye
using a graphical tool. If a particular light curve was clearly characteristic
of periodic stellar variability, the associated position was flagged as
hosting a ``variable'' star. On the other hand, if a light curve exhibited
behaviour characteristic of a microlensing candidate or other transient
event, it was flagged as ``interesting''. The interesting events were further 
scrutinized by examining both the subtracted and unsubtracted images at their 
respective positions and checking for bad pixels, nearby saturated stars, 
bleeding columns etc. Interesting events were closely monitored throughout 
the viewing campaign.
If any event could be clearly identified as a
microlensing candidate or another tyoe of interesting transient event, a MOA Transient
Alert was issued to the microlensing community and posted on the 
World Wide Web.
\footnote{http://www.phys.canterbury.ac.nz/\~{}physib/alert/alert.html}

Early in the observation season, the number of light curves to inspect by eye
was large but this number decreased rapidly as more and more periodic
variables were flagged. As the observation season progressed, around 100--150 
previously undetected objects were found
each time a search for new events was carried out. This number of 
corresponding light curves could easily be scanned by eye and
classified. There is 
merit in being able to pick up events by eye in real time. No assumptions
on the form of the variability of transient events need be made in the 
selection
process. As well as microlensing events, the possibility of detecting
new and unusual transient phenomena arises with such a system.

\subsubsection{Computational Constraints}

\begin{table*}
\centering
\begin{minipage}{140mm}
\caption{Summary of All Detected Transient Events}
\label{allevents}
\begin{tabular}{@{}lcrccll@{}}

\multicolumn{3}{c}{MOA ID} & 
\multicolumn{2}{c}{Coordinates (J2000.0)} & 
\multicolumn{2}{c}{Alert Status} \\

\multicolumn{1}{l}{Field} & 
\multicolumn{1}{c}{CCD}   & 
\multicolumn{1}{c}{ID}    &
\multicolumn{1}{c}{RA}    &
\multicolumn{1}{c}{Dec}   &
\multicolumn{1}{c}{MOA}   & 
\multicolumn{1}{c}{OGLE} \\[10pt]

ngb1  & 2 & 2667 & 17:54:56.68 & $-$29:31:47.5 & 2000--BLG--7  & \\
ngb1  & 2 & 2717 & 17:57:07.91 & $-$29:09:59.3 & 2000--BLG--11 & \\
ngb1  & 3 &  727 & 17:54:29.77 & $-$28:55:59.3 & 2000--BLG--3  & \\
ngb1  & 3 & 2540 & 17:58:20.94 & $-$28:47:48.8 & & \\
ngb1  & 3 & 2548 & 17:55:05.43 & $-$28:50:34.6 & & \\
ngb2  & 2 & 1648 & 18:00:12.36 & $-$29:37:23.9 & & \\
ngb3  & 2 & 1316 & 18:05:09.53 & $-$30:36:06.8 & 2000--BLG--9  & \\
ngb4  & 1 & 2806 & 17:55:33.20 & $-$28:10:17.1 & 2000--BLG--13 & \\
ngb4  & 2 & 2197 & 17:57:20.26 & $-$27:46:21.2 & & \\
ngb4  & 3 &  159 & 17:57:47.20 & $-$27:33:52.8 & & \\
ngb5  & 1 & 1616 & 18:01:30.87 & $-$28:59:26.9 &             & 2000--BUL--30 \\
ngb5  & 1 & 1629 & 17:59:58.11 & $-$28:48:19.1 & & \\
ngb5  & 1 & 1668 & 18:01:44.79 & $-$28:58:03.5 &             & 2000--BUL--48 \\
ngb5  & 1 & 1672 & 18:01:26.81 & $-$28:52:34.7 &             & 2000--BUL--62 \\
ngb5  & 1 & 1673 & 18:01:06.71 & $-$28:52:22.3 &             & 2000--BUL--55 \\
ngb6  & 3 & 1425 & 18:03:54.78 & $-$28:34:58.6 & 2000--BLG--12 & 2000--BUL--64 \\
ngb7  & 3 &  703 & 18:10:55.62 & $-$29:03:54.2 & 2000--BLG--8  & \\
ngb9  & 3 &  841 & 18:10:17.99 & $-$27:31:19.3 & & \\
ngb10 & 1 &  211 & 18:08:07.97 & $-$26:13:14.4 & & \\
ngb11 & 2 & 1011 & 18:11:51.18 & $-$26:26:48.6 & & \\
ngb11 & 2 & 1142 & 18:11:28.31 & $-$26:15:05.8 & 2000--BLG--10 & 2000--BUL--33 \\
ngb11 & 2 & 1146 & 18:10:37.44 & $-$26:20:00.1 & & \\
ngb12 & 2 & 1052 & 18:14:47.42 & $-$25:32:53.6 & & \\
ngb13 & 2 & 1170 & 18:16:56.78 & $-$23:29:50.9 & 2000--BLG--6  & \\

\end{tabular}
\end{minipage}
\end{table*}
The reduction of observational images using image subtraction online 
is demanding on
computer resources such as CPU time and disk space. The data analysis 
environment system described here was implemented on a dedicated 4-processor
Sun Enterprise 450 computer located at the observatory site. Online disk space
comprised 90 GB internal and 300 GB external and was expanded from 
time-to-time. 
Raw images entered the analysis pool shown in Fig.~\ref{roadmap} soon 
after acquisition and
certification by the observing staff. A good night during the 
southern winter observing season yielded up to 6 GB of raw images. In 
this case, the analysis took 24 hours to completely reduce the
night's observations. However, it was not necessary to wait until all current
observations were reduced before searching for new events. 

Thus, despite the considerable demands on computer resources, we were 
successful
in implementing an on-line image subtraction analysis system along
with developing an effective system for issuing alerts of microlensing
and other transient events in progress during 2000.

\subsection{Selected Events and Follow-up Analysis}

\begin{figure}
   \centerline{\psfig{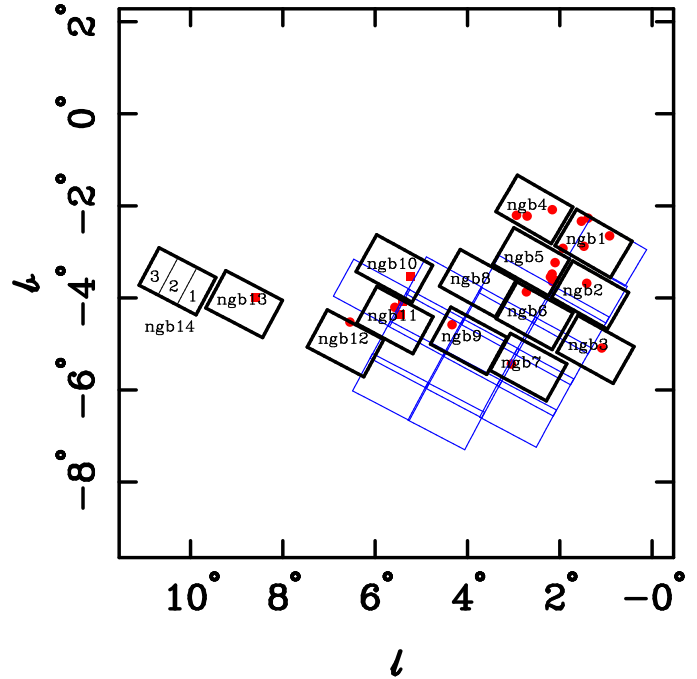}}
   \caption{Locations of the 14 MOA Galactic Bulge fields and positions
	of transient events detected in 2000 shown in Galactic coordinates.
	Candidate microlensing events detected during 2000
	are shown by the dots and the nova-like events are shown by
	the squares. The relative positions of the three CCDs are shown 
        for field ngb14. The MOA fields observed during 1998--99 are also
	shown.}
   \label{gcplot}
\end{figure}


A number of interesting transient events were detected in real time
during the 2000 Galactic Bulge observation season. These are listed
in Table~\ref{allevents}. Their positions are
plotted in Galactic coordinates together with the MOA Bulge fields
in Fig.~\ref{gcplot}. The majority of the events listed are 
probably microlensing events, but the list also includes some nova-like 
events. Nine of the events listed in Table~\ref{allevents} were issued 
as transient alerts.

The online analysis described in the previous sub-section involved a
trade-off between analytical rigour and computing time. A detailed
follow-up analysis was carried out on the events listed in 
Table~\ref{allevents}. For a given
event, 400$\times$400 pixel subframes were extracted from all flat-fielded
observation images in the sequences. An astrometric reference was again
selected from amongst the best quality sub-frames and the geometric
alignment was repeated. Since we were now dealing with small images, we formed
the subtraction reference by combining a selection of the very best 
seeing images. We required that the geometric alignment images selected in 
forming the reference
involved small offsets from the astrometric reference less than 5 pixels. This
ensured any pixel defects are localised to small regions on the combined 
image after geometrically aligning the individual images. 
Furthermore, as with the online
analysis, we imposed the strict requirement that all stamps on the reference
image used to solve for the kernel, be completely free of pixel defects. The
image subtraction process was again carried out on all subframes. The use of
small subframes made it easier to model spatial variations in the kernel
and the differential background using smooth functions.

On a given observation image, the profile on a subtracted image at the
position of a variable star follows the point-spread function on the
corresponding unsubtracted image. In the follow-up analysis the integrated
flux over these profiles was measured using empirical PSF models. For each
event, a high signal to noise PSF was constructed from the reference image
by combining small sub-frames centred on a number of bright, isolated,
and non-saturated stars
near the event position. For each observation image, this reference PSF was
convolved with the corresponding kernel model derived for this image and
evaluated at the event position to obtain a normalized empirical PSF profile
for the given event on the given observation image. This PSF was then rescaled
to fit the observed flux profile on the subtracted image to provide a
measurement of the flux difference. The associated errors were determined in 
a self-consistent manner by applying  the same measurement technique at the 
positions of constant stars in the field (which have flux differences of
zero). The scatter in the flux difference measurements for these stars
determined the frame-to-frame errors.

\subsection{Interpretation of Difference Imaging Photometry}

\begin{figure}
   \centerline{\psfig{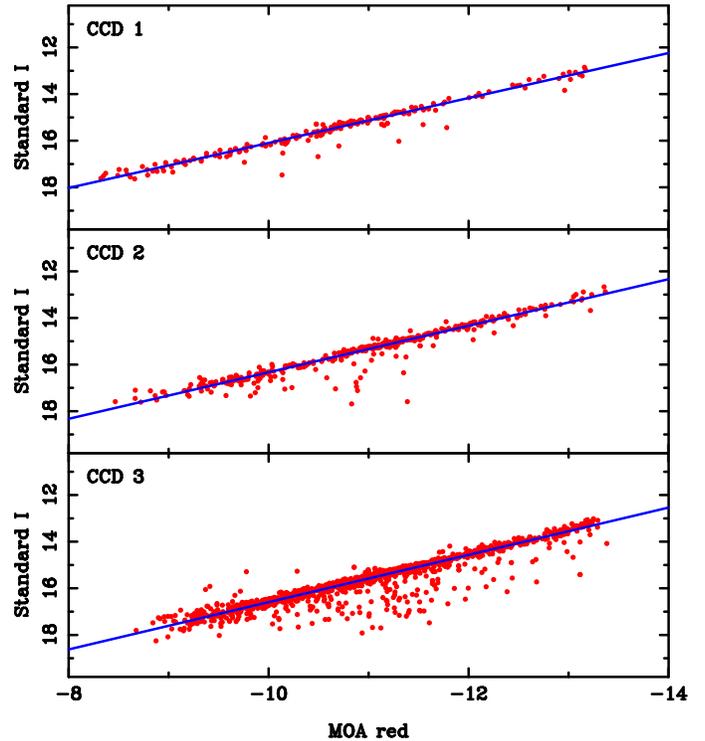}}
   \caption{Calibration of MOA custom red passband measurements to the
	standard I band magnitude scale for the three CCDs of the
        camera \mcII.}
   \label{calib}
\end{figure}

Difference imaging measures flux differences. This is a
relatively unusual concept in astronomical photometry. To convert
``delta flux'' ($\Delta F$) measurements onto a magnitude scale, 
it is necessary to determine a corresponding ``total flux''.
This is given by

\begin{equation}
F = \fr + \Delta F
\end{equation}
where $\fr$ is the flux of the object on the subtraction 
reference image. The reference flux may be measured by applying
a profile fitting program such as Dophot to the reference image.
Alternatively, one can derive the reference flux by fitting
a theoretical profile, such as a microlensing profile, to the 
observed delta-flux light curve. 

There are some situations where it may not be possible to measure the
reference flux, and hence the total flux, for a particular event. For
example, the event may be strongly blended with a close bright star making
a reference flux measurement impossible. In such a situation, classical
profile fitting photometry will not be able to yield a reliable light curve
but difference imaging photometry will at least yield a delta flux light 
curve with each measurement unaffected by blending.

To obtain information in standard passbands,
we used UBVI measurements of selected stars in Baade's Window
provided by OGLE \cite{pac99} that 
are contained in some of the MOA Bulge fields. The Dophot procedure was
applied to the corresponding MOA reference images and the resulting 
star positions were cross-referenced with those of stars with 
UBVI measurements. As seen in Fig.~\ref{calib} we find that measurements 
taken in
the non-standard MOA red passband correlate well with measurements taken 
in the standard I passband. If the total flux of a particular measurement
obtained by image subtraction photometry is known, a corresponding standard
I measurement can be determined by the following relations for the three CCDs of
\mcII

\begin{equation}
 I = \left\{ \begin{array}{ll}
                  25.8016 + 0.9704 m_{\rm red} & \mbox{for CCD 1}\\
                  26.2964 + 0.9970 m_{\rm red} & \mbox{for CCD 2}\\
                  26.9550 + 1.0316 m_{\rm red} & \mbox{for CCD 3}
               \end{array}
       \right.
\end{equation}
where $m_{\rm red}=-2.5\log(F)$ is the instrumental MOA red magnitude
associated with the total flux measurement.

\section{Events Detected During 2000}

\subsection{Microlensing Events}

\begin{figure*}
   \centerline{\psfig{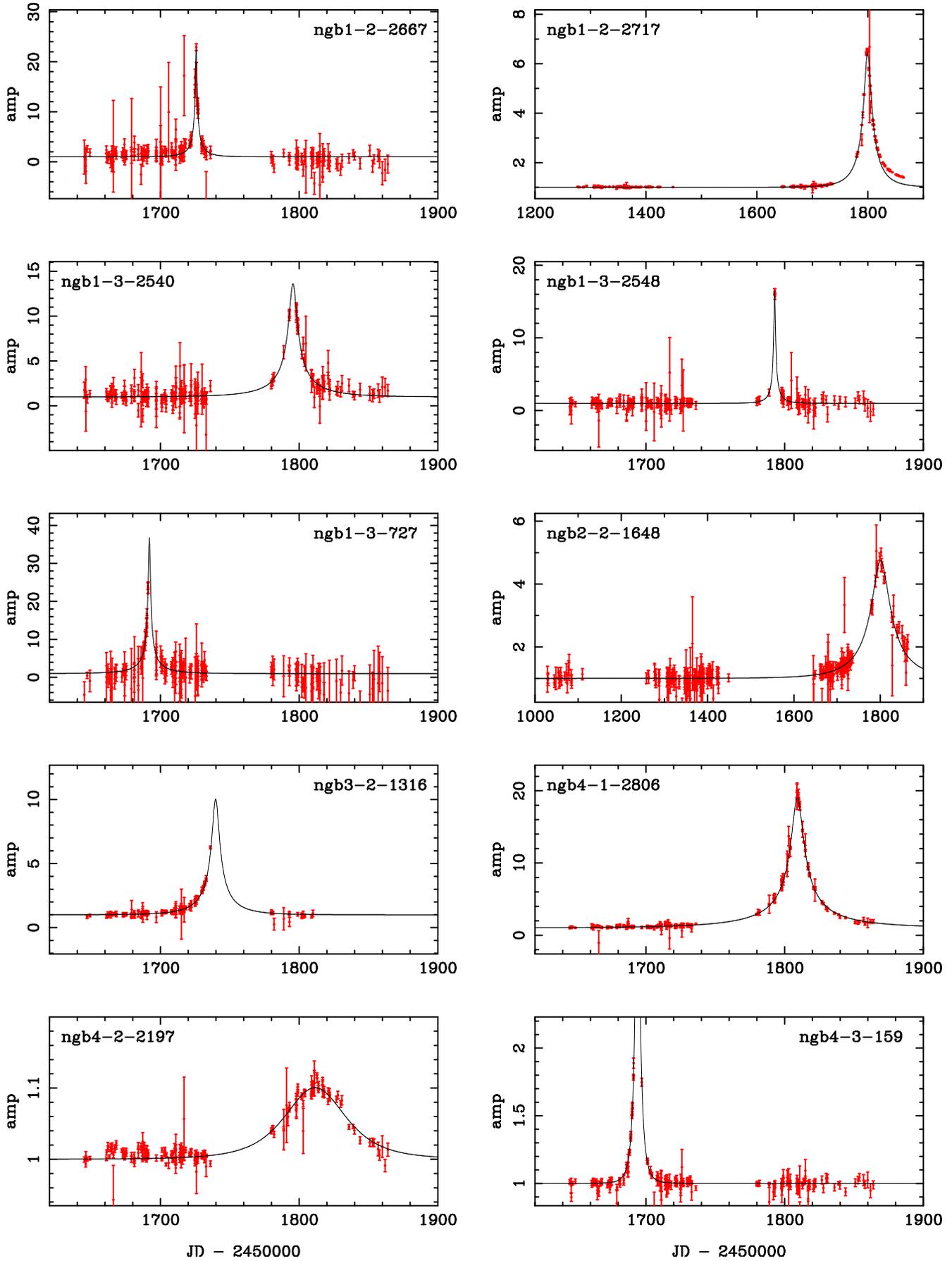}}
   \caption{Candidate microlensing events detected in MOA Galactic Bulge
	fields ngb1, ngb2, ngb3, and ngb4 during 2000.}
   \label{micro1}
\end{figure*}
\begin{figure*}
   \centerline{\psfig{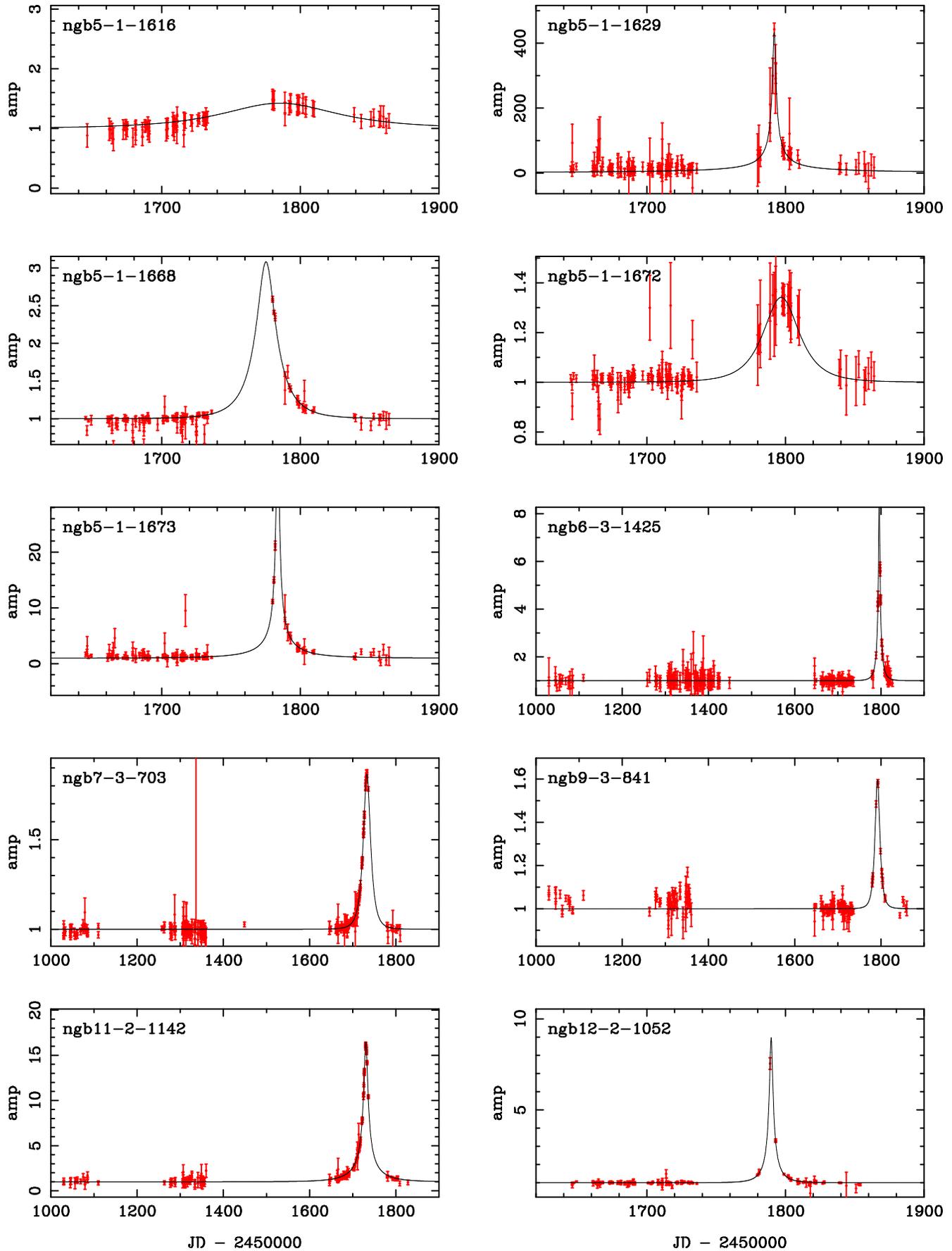}}
   \caption{Candidate microlensing events detected in MOA Galactic Bulge
	fields ngb5, ngb6, ngb7, ngb9, ngb11, and ngb12 during 2000.}
   \label{micro2}
\end{figure*}

\begin{figure}
   \centerline{\psfig{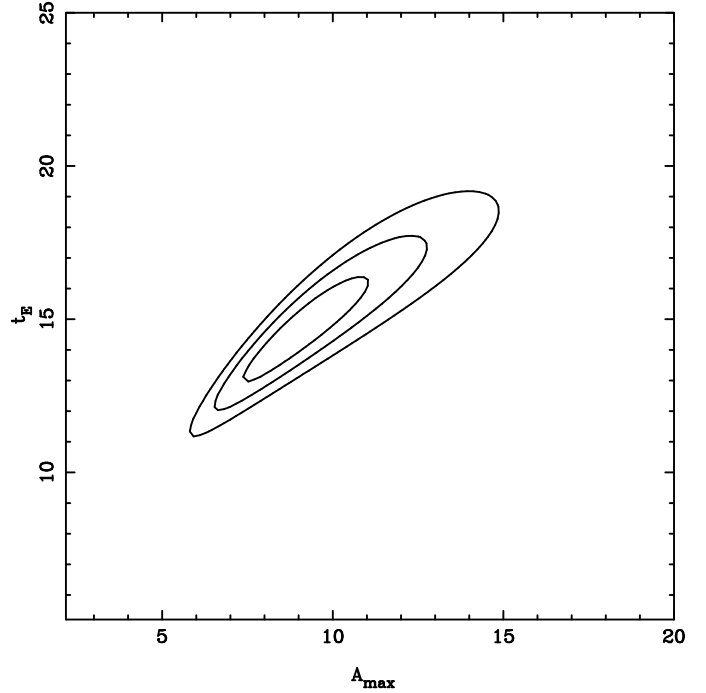}}
   \caption{Contour plot of $\chi^2$ based on fitting a theoretical 
	microlensing profile in ($\amax$, $\that$) parameter space for
	event ngb6--3--1425. The contours correspond to confidence levels
	of 68\%, 95\%, and 99.73\%.}
   \label{contour1}
\end{figure}
\begin{figure}
   \centerline{\psfig{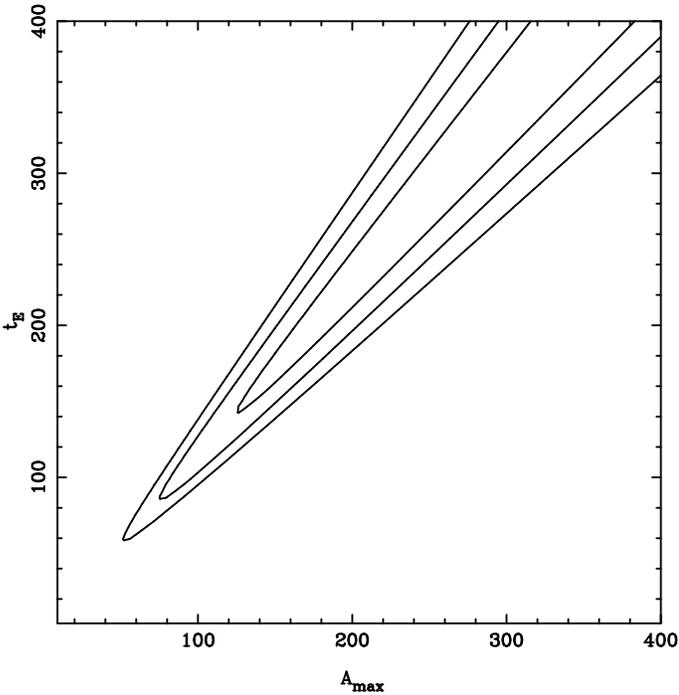}}
   \caption{Contour plot as in Fig.~\ref{contour1} for event ngb5--1--1629.}
   \label{contour2}
\end{figure}

The observed flux difference for a microlensing event is given by
\begin{equation}
   \Delta F = \fb A\left(u(t)\right) - \fr
\label{dflux}
\end{equation}
where $\fb$ is the baseline flux and $\fr$ is the flux on the
subtraction reference image. In general the reference image contains some 
lensed flux and is not equal to the baseline flux.
Even if the source is not visible at baseline, the best seeing images used
to form the subtraction reference image may occur when the source
is lensed. This is true for some of the microlensing events presented here.

If the reference image was formed by combining $N_{\rm ref}$ images taken
at times $\{t_i\} $, the reference flux for a microlensing event is given
by
\begin{equation}
\fr = \frac{\fb}{N_{\rm ref}} \sum_i A(u(t_i))
\end{equation}

The amplification, $A(u)$, expressed in terms of the distance, $u$, of the
lens star from the line-of-sight to the source star expressed in units of
the Einstein radius is given by \cite{pac86}
\begin{equation}
A(u)=\frac{u^2+2}{u\sqrt{u^2+4}}
\label{amp}
\end{equation}
where 
\begin{equation}
   u(t)=\sqrt{\umin^2 + \left(\frac{t-\tmax}{\that}\right)^2}.
\label{impact}
\end{equation}
Here $\umin$ is the minimum value of $u$ during an event, $\that$ is
the characteristic Einstein radius crossing timescale, and $\tmax$ is the
time of maximum amplification. The time $\that$ is equal to $\re/v_{\rm T}$
where $v_{\rm T}$ is the transverse velocity of the lens with respect to the 
line-of-sight to the source.

\begin{table*}
\centering
\begin{minipage}{140mm}
\caption{Fitted Parameters of Microlensing Candidates}
\label{micros-table}
\begin{tabular}{lcrrrrrrrcc}

\multicolumn{3}{c}{MOA ID} & 
\multicolumn{3}{c}{$\amax$} & 
\multicolumn{3}{c}{$\that$ / days} &
\multicolumn{1}{c}{$\tmax$} &
\multicolumn{1}{c}{I magnitude} \\

\multicolumn{1}{l}{Field} & 
\multicolumn{1}{c}{CCD}   & 
\multicolumn{1}{c}{ID}    &
\multicolumn{1}{c}{lower} &
\multicolumn{1}{c}{best}  &
\multicolumn{1}{c}{upper} & 
\multicolumn{1}{c}{lower} &
\multicolumn{1}{c}{best}  &
\multicolumn{1}{c}{upper} &
\multicolumn{1}{c}{JD$-$2450000} &
\multicolumn{1}{c}{baseline} \\[10pt]

ngb1  & 2 & 2667 &  10.7 &  22.8 &     - &   8.0 &  14.8 &     - & 1725.74 & 19.96\\
ngb1  & 2 & 2717 &   5.8 &   6.5 &   7.0 &  40.2 &  43.0 &  46.1 & 1799.62 & 14.16\\
ngb1  & 3 &  727 &  12.8 &  36.8 &     - &  11.0 &  29.1 &     - & 1691.98 & 20.20\\
ngb1  & 3 & 2540 &   5.9 &  13.6 &  60.8 &  24.2 &  40.7 & 140.8 & 1795.54 & 19.24\\
ngb1  & 3 & 2548 &   5.0 &  16.5 & 118.4 &   5.0 &   9.5 &  40.0 & 1792.80 & 19.32\\
ngb2  & 2 & 1648 &   3.5 &   4.8 &   6.6 &  79.2 &  95.7 & 118.4 & 1801.13 & 17.74\\
ngb3  & 2 & 1316 &   3.2 &  10.0 &     - &  11.2 &  26.7 &     - & 1739.76 & 18.03\\
ngb4  & 1 & 2806 &  15.7 &  19.0 &  24.3 &  69.4 &  83.0 & 104.6 & 1809.15 & 17.49\\
ngb4  & 2 & 2197 &   1.0 &   1.1 &   3.3 &  14.0 &  19.8 &  64.0 & 1812.50 & 14.11\\
ngb4  & 3 &  159 &   2.0 &   3.7 &     - &   3.9 &   5.1 &   6.5 & 1693.98 & 15.80\\
ngb5  & 1 & 1616 &   1.3 &   1.4 &     - &  39.0 &  57.2 &     - & 1785.17 & 15.12\\
ngb5  & 1 & 1629 &  68.2 & 434.8 &     - &  84.0 & 507.0 &     - & 1791.86 & 22.74\\
ngb5  & 1 & 1668 &   1.4 &   3.1 &  15.0 &  12.0 &  20.3 &  60.0 & 1775.17 & 15.74\\
ngb5  & 1 & 1672 &   1.1 &   1.3 &   4.3 &   9.0 &  16.6 &  44.0 & 1796.93 & 15.38\\
ngb5  & 1 & 1673 &  27.8 &  49.1 &     - &  29.6 &  42.3 &  72.0 & 1783.65 & 18.11\\
ngb6  & 3 & 1425 &   6.3 &   8.7 &  12.4 &  12.0 &  14.5 &  17.6 & 1796.00 & 17.42\\
ngb7  & 3 &  703 &   1.6 &   1.9 &   2.2 &  13.1 &  15.2 &  17.1 & 1732.87 & 15.03\\
ngb9  & 3 &  841 &   1.2 &   1.6 &   5.0 &   6.0 &   8.8 &  20.0 & 1791.89 & 14.71\\
ngb11 & 2 & 1142 &  13.9 &  16.5 &  20.2 &  63.1 &  74.4 &  90.1 & 1730.31 & 17.65\\
ngb12 & 2 & 1052 &   6.3 &   9.0 &  31.1 &   8.8 &  10.9 &  14.6 & 1789.68 & 16.80\\

\end{tabular}
\end{minipage}
\end{table*}

We fitted the theoretical single lens light curve given by 
Eqns~\ref{dflux}--\ref{impact} to the data for candidate 
microlensing events using four parameters $\fb$,  $\umin$, $\that$,
and $\tmax$. The light curves are shown in Figs.~\ref{micro1}--\ref{micro2}
where the $\Delta F$ data points have been converted to amplifications using 
the fitted parameters. We note that for some of the events, the baseline flux
is below the observational threshold. However, we could still derive a
best fit value for $\fb$. This is essentially an extrapolation of
the light curve from where the event is visible down to the baseline level.
In these events this value for $\fb$ is used to calculate amplifications in
plotting the light curves in Figs.~\ref{micro1}--\ref{micro2}. We address
the issue of measuring microlensing parameters for these types of events
in detail below.

The microlensing parameters $\that$ and the peak amplification $\amax$ 
(calculated directly from $\umin$ using Eqn.~\ref{amp}) are the interesting 
physical quantities in a microlensing event. Both quantities can be measured
for microlensing events with good sampling from the baseline to the peak. However,
for high magnification events sampled only around the times of peak amplification,
it is possible only to measure the quantity $t_{\rm eff}=\that\umin$ which is 
degenerate in the two parameters of interest \cite{go}. We investigated the extent to
which these parameters can be constrained by mapping $\chisq$ in the 
two dimensional parameter space corresponding to $\amax$ and $\that$. 
For each light curve, we fitted Eqns~\ref{dflux}--\ref{impact} for a sample
of fixed $(\amax, \that)$ pairs with the remaining parameters, $\fb$ and $\tmax$,
allowed to float.

In Fig.~\ref{contour1} we show a contour map of 
$\chisq$ in $(\amax, \that)$ space for the event ngb6--3--1425.
The contours connect equal values of $\chisq_{\rm min}+\Delta\chisq$ for
$\Delta\chisq=(2.3, 6.2, 11.8)$. These formally correspond to confidence limits
of 68\%, 95\%, and 99.73\% respectively for the simultaneous estimation
of two interesting parameters \cite{av}. Event ngb6--3--1425 can be described 
as a ``classical''
microlensing event where the source was resolved and visible at baseline and
the light curve was well sampled both at its baseline and during its 
lensing phase. In this case both $\amax$ and $\that$ are well constrained.

In contrast the parameter space of the event ngb5--1--1629 was not as well
constrained as can be seen in Fig.~\ref{contour2}. A close inspection of 
the baseline
images for this event did not show any visible object at the position of
the source while it was clearly visible during amplification. This event
is clearly characteristic of one which is amplified from below the 
observational threshold to above the threshold. Such events are only
observable around the times of high magnification, and only lower
limits can be inferred for $\amax$ and $\that$. It is nevertheless 
noteworthy that one can infer that this event had a peak amplification 
well in excess of 70.

Alcock et al. \shortcite{al00} proposed classifying microlensing events 
found by DIA as
either classical or ``pixel lensing'' events following the definition by
Gould \shortcite{go} of pixel lensing events as microlensing of 
unresolved sources. A
source may be unresolved if it is strongly blended with nearby stars or if it 
is too faint to be observed at baseline. Since DIA is unaffected by blending
we suggest here a classification based just on amplification. Events would
then fall into two types: those amplified from below the observation
threshold and those amplified from above. For those events whose baselines
are below the detection and hence measurement threshold, it is only
possible to obtain lower limits on $\amax$ and $\that$. 
We suggest that in DIA it is important to be able to distinguish 
between the two types of events as this affects the extent to which the
microlensing parameters can be constrained.

In Table~\ref{micros-table} we summarize the results of this 
analysis for all 20 microlensing
events showing best fit values for $\amax$ and $\that$ along with 95\%
lower and upper limits where possible. The fitted baselines fluxes were
converted to I magnitudes using the calibration information described in
Section 3.4. The parameters of one event, 
ngb3--2--1316, are poorly constrained as this event did not receive 
adequate temporal coverage. 

Of the 20 events listed in Table~\ref{micros-table}, half of them 
have best fit peak amplifications around 10 or more, and six have
lower limits of $\amax$ greater than 10. This is a large fraction of
high magnification events when compared with the corresponding fractions
from the MACHO and OGLE surveys. Since these groups use profile fitting 
photometry in their on-line analysis, their measurements are in general 
affected by blending and therefore the derived values of $\amax$ tend to be 
underestimated. More importantly, DIA is sensitive to microlensing events 
whose source stars are too faint to be detected at baseline.

The large fraction of detected events with high magnification has useful
implications for studies of extra-solar planets, low mass stars and stellar 
atmospheres, as noted in the introduction.

\subsection{MOA--2000--BLG--11: A Microlensing Parallax Event}

\begin{figure}
   \centerline{\psfig{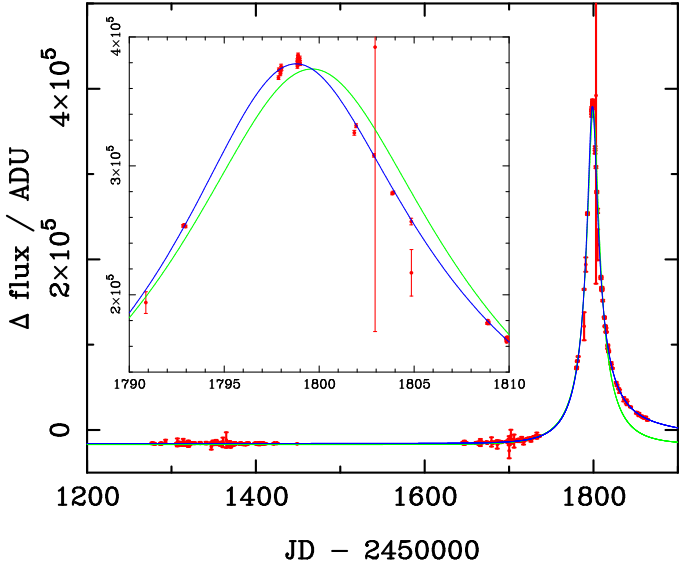}}
   \caption{Delta-flux light curve of MOA--2000--BLG--11 with standard 
	and parallax microlensing fits.}
   \label{parallax}
\end{figure}

\begin{table}
\centering
\caption{MOA-2000--BLG--11: Standard and parallax microlensing fit parameters}
\label{parallax-table}
\begin{tabular}{lll}

Parameter      & Standard    & Parallax \\[10pt]

$I_{\rm base}$       & 14.16       & 14.78     \\ 
$t_0$                & JD2451799.62     & JD2451837.62   \\
$\that$              & 43.04 days      & 69.69 days    \\
$\umin$              & 0.157       & 0.224     \\
$\vbar$              &   -         & 42.52 km s$^{-1}$ \\
$\theta$             &   -         & $-$76$\fdg$20 \\
$\chisq$/dof         & 18027.6/401 & 1071.6/399 \\

\end{tabular}
\end{table}

The light curve of event ngb1--2--2717, which was alerted as 
MOA--2000--BLG--11, is not well described by the standard microlensing 
profile given by Eqns.~7--8. 
The light curve exhibits an asymmetric profile which is characteristic of 
parallax microlensing. Here, a distortion arises from the circularly
accelerating motion of the Earth around the Sun. 
When including this motion, the distance between the
lens star and the source star (given by Eqn.~8 for the constant velocity case) 
contains two extra parameters: the angle, $\theta$, between the transverse
motion of the lens and the transverse speed, $\vbar$, projected onto the 
solar position. The full mathematical details can be found in 
Alcock et al. \shortcite{al95}, Mao \shortcite{mao}, 
and Dominik \shortcite{do}.

The light curve of MOA--2000--BLG--11 is shown in Fig.~\ref{parallax} 
along with the best fit
standard constant velocity and parallax microlensing models.
The derived parameters are given in Table~\ref{parallax-table}. 
At 75.1 days, the duration of
MOA-2000--BLG--11 is rather short compared with the durations for
previously detected parallax events of 
111.3 days for the first parallax event found by the MACHO group \cite{al95}, 
118.1 days for OGLE--99--CAR--1 \cite{mao}, and 
156.4 days for OGLE--00--BUL--43 \cite{sos}. 

\begin{figure*}
   \centerline{\psfig{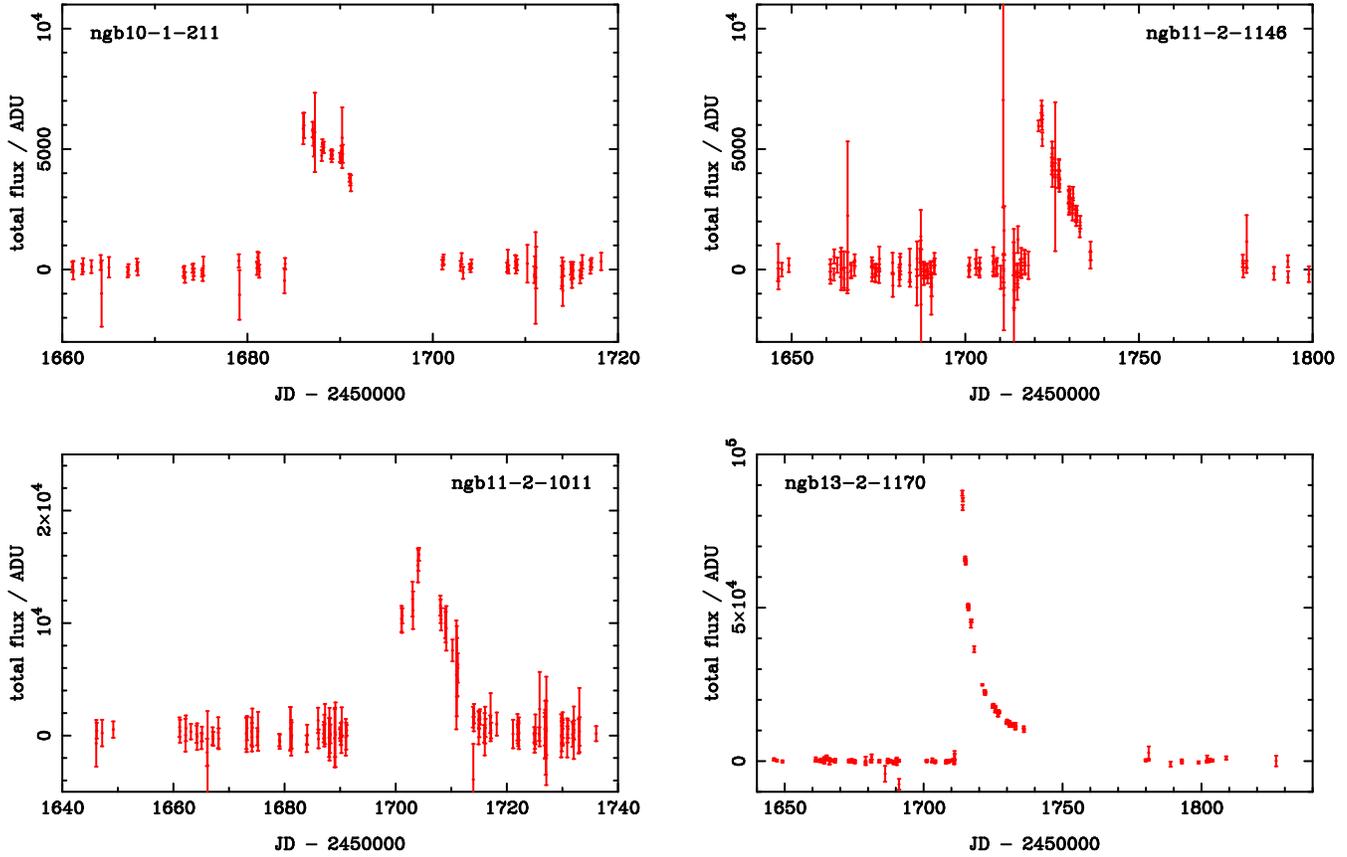}}
   \caption{Light curves of nova-like events detected during 2000.}
   \label{novae}
\end{figure*}

Measurements of microlensing parallax events partially resolve the 
degeneracy in the mass, lens distance, and transverse velocity contained
in the standard microlensing equation. Assuming negligible velocity 
dispersion in the disk and the bulge, the transverse speed projected onto
the solar position can be written as

\begin{equation}
\vbar = \frac{220 x}{1-x} {\rm km s}^{-1}
\end{equation}
where $x=D_l/D_s$ with $D_l$ being the distance to the lens and 
$D_s$ the distance to the source star. Following Alcock et al. 
\shortcite{al95} the mass of the lens, $M_l$, can written as

\begin{equation}
M_l = \frac{1-x}{x} \frac{\vbar^2 \that^2 c^2}{4GD_s}.
\end{equation}
Substituting the parameters obtained from our parallax fit and 
assuming $D_s=8$ kpc we obtain $D_l\sim1.4$ kpc and $M_l\sim0.3 M_{\sun}$
for the lens star. It would then contribute very little light through the 
course of the event. It should be noted that, provided the lens star is 
constant, the contribution of any flux from the lens star is canceled
out in delta-flux measurements.

\subsection{Nova-like Events}

\begin{table}
\centering
\caption{Summary of the characteristics of nova-like events detected in 2000.
The column, $\Delta I$, denotes the change in I magnitude during the rise of
the event from its quiescent state to its high state.}
\label{novae-data}
\begin{tabular}{@{}cccclc@{}}

\multicolumn{3}{c}{MOA ID} &

\multicolumn{1}{c}{peak I} & 
\multicolumn{1}{c}{$\Delta I$} & 
\multicolumn{1}{c}{$t_{1/2}$}\\

\multicolumn{1}{l}{Field}  & 
\multicolumn{1}{c}{CD}    & 
\multicolumn{1}{r}{ID}     & & &days \\[10pt]

ngb10 & 1 &  211 & 16.64 & $>$2.4 & 5 \\
ngb11 & 2 & 1011 & 15.78 & $>$3.3 & 3 \\
ngb11 & 2 & 1146 & 16.75 & $>$2.3 & 7 \\
ngb13 & 2 & 1211 & 13.95 & $>$5   & 4 \\

\end{tabular}
\end{table}

Some events with light curves showing nova-like profiles were found. We 
present them in Fig.~\ref{novae} and Table~\ref{novae-data} to illustrate the 
potential for detecting new types of objects.
None of the events was
visible in the MOA images during their times of minimum light. 
Event ngb13--2--1170 (alerted as MOA--2000--BLG--6) is clearly 
characteristic of a dwarf nova. This event
rose by more than five magnitudes in less than three days peaking at $I=13.95$.
Event ngb11--2--1146 also appears to be the tip of a dwarf nova decay just
above the detection limit. Event ngb10--1--211 is also nova-like but the
decay is not as smooth as the other events.
At the position of event ngb11--2--1011, the 
MACHO database contains an object whose corresponding light curve undergoes 
flaring episodes about once per year (D. Bennett, private communication).

\subsection{Asteroids}

As well as measurements on variable stars, difference imaging analysis can
be employed to study moving objects. In Fig.~\ref{asteroid} 
we show subimages of
three exposures of MOA field ngb1 on CCD number 1 taking during one night.
The subimages include the original unsubtracted images along with their
corresponding subtracted images. The exposures were taken over a time 
interval of approximately
1.6 hours. An object moving across the field of view can
clearly be seen. The positions and magnitudes of this object measured
on each image are listed in Table~\ref{asteroid-table}. A crosscheck 
with {\it Guide~7.0} revealed asteroid (3680) Sasha at these 
positions and times.


The associated transverse velocity across the sky for this object is 
$19.2 \arcsec$ hour$^{-1}$. Frequent sampling of the fields
are required if one is to track the motion of such fast moving objects 
during a night. Sampling the fields once per night would merely result in 
spike events at asteroid positions. The MOA data base now contains 
subtracted images derived from
a large number of nights where the sampling rate was three or more times
per night. We are currently exploring ways of exploiting this data base
to study asteroids.

\begin{table}
\centering
\caption{Astrometry and Photometry of Asteroid}
\label{asteroid-table}
\begin{tabular}{@{}cccc@{}}

& \multicolumn{2}{c}{Coordinates (J2000.0)}\\

JD-2450000 & RA & DEC & I magnitude\\[10pt]

1675.066760 & 17:56:23.33 & $-$29:54:51.4 & 15.33 \\
1675.133438 & 17:56:21.28 & $-$29:55:06.5 & 15.19 \\
1675.197859 & 17:56:19.28 & $-$29:55:21.0 & 15.28 \\

\end{tabular}
\end{table}

\section{Summary}

It has been demonstrated that real-time difference imaging of an extensive 
database is possible. The procedure employed by MOA is a database management 
system rather than a linear reduction pipeline. A variety of levels of
automation are possible but the final selection of events is by human
intervention. 

To minimize hardware constraints, two reduction procedures
were developed, one for online analysis of large numbers of large images, and
one for follow-up, accurate analyses of selected events. As our on-site
computing power expands, we plan to transfer elements of our follow-up
procedure to the online analysis. We also plan to increase the conversion 
rate of internally monitored candidate events to actual alerts issued to
the microlensing community.

We have demonstrated that a high fraction of events detected by difference
imaging have high magnification. This should have useful implications for
studies of extra-solar planets and faint stars provided the peaks of 
future events are monitored photometrically and/or spectroscopically by
the microlensing and other astronomical communities.

Our observational strategy has been to monitor a moderate area of the
Galactic Bulge, $17\deg^2$, fairly frequently, i.e. up to six times per night.
This is a compromise between survey area and sampling rate that we plan
to continue. It has enabled the detection of some unusual types of events, 
including nova-like transients and an asteroid, as well as microlensing
events of high magnification.

\section*{Acknowledgments}

We thank Dave Bennett and Andy Becker of the MACHO Collaboration and
Andrzej Udalski of the OGLE Collaboration for their invaluable assistance in
identifying microlensing events through cross checking with their
respective databases. We also thank Christophe Alard for pointing out a
subtle issue involved in fitting microlensing curves to difference imaging 
photometry. The Marsden Fund of New Zealand and the Ministry
of Education, Science, Sports, and Culture of Japan are thanked for the
financial support that made this work possible.

\begin{figure}
   \caption{Superposition of sub-regions of 3 subtracted images taken
	on one night showing an asteroid trail.}
   \label{asteroid}
\end{figure}


\appendix

\bsp

\label{lastpage}

\end{document}